\documentclass[article]{JHEP3}
\usepackage{amsmath,amssymb}
\usepackage{cite}
\usepackage{float}
\usepackage{subcaption}
\usepackage{amsfonts}
\usepackage[utf8]{inputenc}
\usepackage{graphicx}

\allowdisplaybreaks

\newcommand{\be}{\begin{equation}}
\newcommand{\ee}{\end{equation}}
\newcommand{\bea}{\begin{eqnarray}}
\newcommand{\eea}{\end{eqnarray}}
\newcommand{\bean}{\begin{eqnarray*}}
\newcommand{\eean}{\end{eqnarray*}}

\def\beq{\begin{equation}}

\def\eeq{\end{equation}}
\def\R{\mathcal{R}}

\def\Re{\mathop{\rm Re}}
\def\Im{\mathop{\rm Im}}
\relax

\def\a{\alpha'}

\title{The isospectrality of asymptotic quasinormal modes of large Gauss-Bonnet $d$-dimensional black holes}

\author{Filipe Moura$^a$ and Jo\~ao Rodrigues$^b$
\\
\\
$^{a}$
Departamento de Matem\'atica, Escola de Tecnologias e Arquitetura, \\ ISCTE - Instituto Universit\'ario de Lisboa \\ and Instituto de Telecomunica\c c\~oes,
\\Av. das For\c cas Armadas, 1649-026 Lisboa, Portugal\\
\email{fmoura@lx.it.pt}
\\
\\
$^{b}$
Centro de An\'alise Matem\'atica, Geometria e Sistemas Din\^amicos,\\ Departamento de Matem\'atica,\\ Instituto Superior T\'ecnico,\\
Av. Rovisco Pais, 1049-001 Lisboa, Portugal\\
\email{joao.carlos.rodrigues@tecnico.ulisboa.pt}
}

\abstract{We compute the quasinormal frequencies of $d$-dimensional large spherically symmetric black holes with Gauss-Bonnet corrections in the
highly damped regime. We solve perturbatively the master differential equation and we compute the monodromies of the master perturbation variable
(analytically continued to the complex plane) in different contours, in order to obtain the quasinormal mode spectra. We consider tensorial,
vectorial and scalar gravitational perturbations, obtaining the same frequencies for the three cases like in Einstein gravity. We also separately
perform the same calculation for test scalar fields.
}

\begin{document}



\vfill

\eject

\section{Introduction}
\noindent

Gravitational wave detectors can directly measure quasinormal ringing frequencies, which carry unique information about parameters of the black hole in the ringdown phase resulting from a binary black hole coalescence \cite{Dreyer:2003bv}. Indeed, the spectrum of black hole quasinormal modes (QNMs) does not depend on what drives the perturbations: it is completely determined by intrinsic physical quantities of the black hole such as mass, charge or spin, and eventually (beyond Einstein gravity) some other parameters of the theory. This feature turns QNMs into preferential probes for testing theories of gravity beyond Einstein, since the ringing frequencies represent a universal part of the gravitational wave signals (for a recent review of this subject see \cite{Franchini:2023eda}). With the advent of gravitational wave astronomy, therefore, interest in the study of black hole QNMs has raised, as their measurement can provide a test to modified gravity theories.

The calculation of the spectra of black hole quasinormal modes has therefore become a very active field of research; for some comprehensive reviews see \cite{Berti:2009kk,Konoplya:2011qq}. Recent works computing QNMs of black holes in different modified gravity theories are \cite{Blazquez-Salcedo:2016enn, Cano:2020cao,Konoplya:2020bxa,Pierini:2021jxd,Anacleto:2021qoe}. Among the most common modifications of gravity are the introduction of extra dimensions and of higher derivative corrections. Both these modifications are required in string theories.

Black hole quasinormal modes are associated to perturbations that can be related either to the black hole metric (gravitational perturbations) or to external fields. They are solutions of a master differential equation (whose form in general depends on the action/black hole solution one is considering) with a potential (depending on the kind of perturbation/external field). Most of the times, QNMs have to be computed numerically. Nonetheless, different analytical methods have been developed in order to compute QNMs in some limiting cases, in four and higher dimensions. Two of such cases are the eikonal limit and the asymptotic (highly damped) limit.

Concretely, the master equation and the corresponding higher-derivative corrected potential have been obtained for tensorial perturbations of $d$-dimensional black holes with leading $\a$ corrections in string theory \cite{Moura:2006pz,Moura:2012fq}. Using this result, we have computed the corresponding quasinormal modes (and also for test scalar fields) in the eikonal limit \cite{Moura:2021eln} and in the asymptotic highly damped limit \cite{Moura:2021nuh}.

The higher-derivative corrected potential has also been obtained for tensorial, vectorial and scalar perturbations of $d$-dimensional black holes with Gauss-Bonnet corrections. The corresponding quasinormal modes in the eikonal limit have been computed in \cite{Konoplya:2017wot}. In this article we compute these quasinormal modes in the asymptotic highly damped limit, like we did in \cite{Moura:2021nuh} in string theory. The calculation of these  quasinormal modes in this regime has been addressed numerically in \cite{Daghigh:2006xg}.

Quasinormal modes associated to different kinds of gravitational perturbations have different spectra in general, but they share the same spectrum in Einstein gravity in $d$ dimensions in some limiting cases. One of such cases is the eikonal limit. Such isospectrality is observed in that limit for spherically symmetric $d$ dimensional black holes in Einstein gravity, but not any longer in the presence of perturbative Gauss-Bonnet corrections. As shown in \cite{Konoplya:2017wot}, the leading terms are equal, but the first order corrections to the real parts of quasinormal frequencies are distinct for the different kinds of gravitational perturbations; the imaginary parts of the same quasinormal frequencies are still equal to first order in the Gauss-Bonnet parameter, but in the same article they have been shown to be distinct to second order in such parameter.

Another limiting case in which there is an isospectrality of quasinormal modes for different kinds of perturbations in Einstein gravity in $d$ dimensions is the asymptotic (highly damped) limit. This is the limit we will address in this work, for all different kinds of perturbations of spherically symmetric black holes, in the presence of perturbative Gauss-Bonnet corrections, hoping to clarify if the isospectrality of quasinormal modes is preserved or not with such corrections. We will also consider quasinormal modes of massless scalar test field in the background of these black holes. This way, for tensorial perturbations and test scalar fields, we will be able to compare the results for the two theories: Einstein-Gauss-Bonnet theory (in which the only modification to $d$ dimensional Einstein gravity is the introduction of higher derivative corrections) and string theory (in which other fields are necessarily present, namely the dilaton).

The article is organized as follows. In section \ref{sstp} we review a $d$-dimensional spherically symmetric black hole solution in Einstein-Gauss-Bonnet gravity and its large black hole limit. In section \ref{ssbhgp} we review the gravitational perturbation theory of these black holes, writing down the master equation and the respective potentials corresponding to these perturbations and also to test scalar fields. Given this information, in section \ref{qnm} we compute the quasinormal spectrum in the highly damped limit corresponding to these perturbations and fields using the monodromy method. We conclude by discussing our results.

\section{Einstein-Gauss-Bonnet black holes}
\label{sstp}

\subsection{The Einstein-Gauss-Bonnet action}
\noindent

In $d$--dimensional Gauss-Bonnet gravity ($d>4$) one considers the following action with higher derivative corrections:
\be \label{eef} \frac{1}{16 \pi G_d} \int_\mathcal{M} \sqrt{-g} \left( \R + \alpha (\R_{mnpq}\R^{mnpq} - 4\R_{mn}\R^{mn}+ \R^2) \right) \mbox{d}^dx,
\ee
where $\mathcal{M}$ is our $d$-dimensional space time manifold, $G_d$ is the generalized gravitational constant and $\alpha$ is a positive coupling constant.

In the context of superstring theories, this action (also always coupled to a dilaton $\phi$ and eventually to other string fields) represents an $\a$ correction to Einstein gravity. The whole set of corrections is expressed in a perturbative expansion in the inverse string tension $\a$, of which the higher order Gauss-Bonnet term in (\ref{eef}) is the leading correction term. From this point of view, (\ref{eef}) (or its equivalent in string theory) is an effective action. The parameter $\alpha$ in (\ref{eef}) is to be identified (up to a numerical constant) with the inverse string tension $\a$. Besides $\alpha$ (or $\a$), in string theory there is also a dilaton term multiplying the Gauss-Bonnet correction, since $e^\phi$ represents the string coupling, $\phi$ being the dilaton field.

In Gauss-Bonnet gravity one takes a different point of view, considering (\ref{eef}) as a complete (and not effective) action and computing exact solutions to its field equations (and not just solutions that are perturbative in $\alpha$; in particular, not assuming that $\alpha$ is necessarily a small parameter). This is one of the two very important conceptual differences between Gauss-Bonnet gravity and superstring gravity, the other one being the presence/absence of the dilaton.

\subsection{A spherically symmetric black hole solution}
\noindent

A general static spherically symmetric metric in $d\geq4$ dimensions can always be written depending only on a metric function $f(r)$:
\be \label{schwarz}
ds^2 = -f(r)\ dt^2  + f^{-1}(r)\ dr^2 + r^2 d\Omega^2_{d-2}.
\ee
$d\,\Omega^2_{d-2}=\sum_{i=1}^{d-2} \prod_{j=1}^{i-1} \sin^2 \theta_j d\,\theta_i^2$ is the canonical metric tensor field of the unit ($d$-2)-sphere.

In references \cite{Boulware:1985wk,Wiltshire:1988uq,Wheeler:1985nh,Wheeler:1985qd}, black hole solutions of the field equations from (\ref{eef}) are found and discussed. We are interested in a spherically symmetric solution of the form (\ref{schwarz}), with
\begin{equation} \label{bw}
    f(r) = 1 + \frac{r^2}{\alpha(d-3)(d-4)}(1-q(r)) \hspace{10pt}, \hspace{10pt}q(r) = \sqrt{1 + \frac{4 \alpha (d-3) (d-4) \mu}{(d-2)r^{d-1}}}.
\end{equation}
The parameter $\mu$ is related to the black hole mass through
$$M= \frac{\Omega_{d-2}}{8 \pi G_d} \mu, \, \Omega_{d-2}=\frac{2 \pi^{\frac{d-1}{2}}}{\Gamma\left(\frac{d-1}{2}\right)}.$$
We introduce the parameter $R_0$ which we define by
\be
\mu = \frac{(d-2)}{2} R_0^{d-3} \hspace{10pt} \Leftrightarrow \hspace{10pt}
R_0 = \left(\frac{2\mu}{d-2}\right)^{\frac{1}{d-3}}. \label{r0}
\ee
$R_0$ would correspond to the horizon radius of the Tangherlini solution, in the absence of Gauss-Bonnet corrections (i.e. setting $\alpha=0$). For the solution we are considering, this parameter has no physical meaning; we introduce it for convenience, and because it is simply related to $\mu$. The true horizon radius $R_H$ of the black hole (\ref{bw}) is related to $R_0$ through
\be
R_0^{d-3} = R_H^{d-3} \left(1 + \alpha \frac{(d-3)(d-4)}{2 R_H^2}\right). \label{mu}
\ee

\subsection{The perturbative large black hole limit}
\noindent

In this article, we take the limit of a small coupling constant $\alpha$. In this limit one is allowed to take a perturbative expansion in $\alpha$. This procedure is similar to the one in string theory, where one also takes a perturbative expansion in $\a$. But the situation there is different, as in string theory one must also consider other fields. Most of these fields can be set to $0$, but not the dilaton, whose couplings imply that it is nonzero and dynamical. Because of that, even in the perturbative small $\alpha$ limit string theory is different from the Einstein-Gauss-Bonnet theory we are considering, where there isn't any dilaton.

More concretely, concerning the black hole solution (\ref{bw}) in which we are interested, defining the dimensionless parameter
\begin{equation}
\lambda' = \frac{\alpha}{R_0^2}, \label{lambda}
\end{equation}
this limit corresponds to the condition $\lambda' \ll 1$ and, from (\ref{mu}), it is equivalent to having $\alpha \ll R_H^2$ i.e. the limit of large black holes. We take the perturbative expansion in $f$ and consider only terms to first order in
$\lambda'$ as
\begin{equation}
    f(r) =  1 - \left(\frac{R_0}{r}\right)^{d-3} + \lambda' \frac{(d-3)(d-4)}{2} \left(\frac{R_0}{r}\right)^{2d-4}.
    \label{1}
\end{equation}
In order to simplify the notation we define the functions
\begin{equation}
 f_0(r) = 1 - \left(\frac{R_0}{r}\right)^{d-3} \hspace{10pt} , \hspace{10pt}  \delta f (r) = \frac{(d-3)(d-4)}{2}
      \left(\frac{R_0}{r}\right)^{2d-4} \frac{1}{1 - \left(\frac{R_0}{r}\right)^{d-3}}, \label{deltf}
\end{equation}
and rewrite (\ref{1}) as
\begin{equation}
    f(r) = f_0(r)(1 + \lambda' \delta f(r)).
    \label{fpert}
\end{equation}
We can also consider a perturbative expansion in $\lambda'$ in obtaining the physical quantities related to the black hole (\ref{bw}) in the limit (\ref{fpert}). The horizon radius $R_H$ can be obtained from taking such expansion in (\ref{mu}), with the result
\begin{equation}
R_H = R_0 \left(1 -\frac{d-4}{2} \lambda'\right). \label{rh}
\end{equation}

The temperature of a spherically symmetric black hole of the form (\ref{schwarz}), like (\ref{bw}), is given by $T_H = \frac{f'(R_H)}{4\pi}.$
In the large black hole perturbative limit we are considering, to first order in $\lambda'$, this temperature reads
\begin{equation}
    T_H = \frac{d-3}{4\pi R_0}\left(1- \lambda' \frac{(d-4)(d-2)}{2}\right). \label{temp}
\end{equation}
We see that, even in this perturbative limit, the black hole solution (\ref{bw}) is different and has distinct properties from the $d-$dimensional stringy solutions with quadratic curvature corrections obtained in \cite{cmp89,Moura:2009it}.

\section{Gravitational perturbations in the perturbative limit}
\label{ssbhgp}
\noindent

General tensors of rank at least 2 on the $(d-2)$-sphere $\mathbb{S}^{d-2}$ can be uniquely decomposed in their tensorial, vectorial and scalar components. That is the case of general perturbations $h_{\mu\nu}=\delta g_{\mu\nu}$ of a $d-$dimensional spherically symmetric metric like (\ref{schwarz}). For such metric we have then scalar, vectorial and (for $d>4$) tensorial gravitational perturbations.

Each type of perturbation is described in terms of master variables $\Psi_a(r,t)$ (the subscript $a$ indicates the kind of perturbation). In Einstein gravity, each master variable obeys a second order differential equation (``master equation''):
\be
\frac{\partial^2 \Psi_a}{\partial x^2} - \frac{\partial^2 \Psi_a }{\partial t^2} = V_a \left[ f(r)\right] \Psi_a.
\label{pot}
\ee
This ``master equation'' is given in terms of the tortoise coordinate $x$ for the metric (\ref{schwarz}) defined by
\be
dx = \frac{dr}{f(r)}, \label{tort}
\ee
and of a potential $V_a \left[ f(r) \right]$ that depends on the kind of perturbation one considers \cite{ik03a}: $V_{\textsf{S}}, V_{\textsf{V}}, V_{\textsf{T}}$. We assume the time dependence of the master variables to be of the form
\be
\Psi(x,t) = e^{i\omega t} \psi(x), \label{time}
\ee
such that $\frac{\partial\Psi}{\partial t} = i\omega \Psi$ (for simplicity we drop the subscript $a$). In this way the master equation (\ref{pot}) may be written in Schr\"odinger form, for a generic potential $V,$ as
\be
\frac{d^2 \psi}{d\, x^2} + \omega^2 \psi = V \left[ f(r) \right] \psi.
\label{potential0}
\ee

In the presence of higher derivative corrections in the lagrangian, one still has spherically symmetric black holes of the form (\ref{schwarz}) and the same type of gravitational perturbations. The master equation obeyed by each perturbation variable may change, though: generically, it may become a higher order differential equation. But for static black holes in $d-$dimensional Lovelock gravity of arbitrary order it has been shown \cite{Takahashi:2009xh,Takahashi:2010ye} that, perturbing the field equations, one also obtains for each perturbation variable a second order master equation like (\ref{potential0}).

Concretely, the master equation and the corresponding higher-derivative corrected potential have been obtained for tensorial perturbations of black holes with leading $\a$ corrections in string theory \cite{Moura:2006pz,Moura:2012fq}.

Gravitational perturbations for this black hole solution have been studied and the higher-derivative corrected potential has also been obtained for tensorial \cite{Dotti:2005sq}, vectorial and scalar \cite{Gleiser:2005ra} perturbations of $d$-dimensional black holes with Gauss-Bonnet corrections. The stability of this solution under such perturbations has also been studied in \cite{Konoplya:2008ix}. These are the cases we will address in this work in order to compute the quasinormal modes. Therefore we will now review the master equations and potentials associated with each of these cases, corresponding to perturbing the solution (\ref{bw}) of the field equations from (\ref{eef}). But first we consider the simpler case of a massless scalar test field.

\subsection{Massless scalar test field}
\noindent

The field equation of a minimally coupled massless scalar test field in the background of a spherically symmetric $d$-dimensional black hole is of
the form of the Schrödinger like master equation (\ref{potential0}), with the minimal effective potential given by
\begin{equation}
     V_{\textsf{M}}(r) = f(r)\left(\frac{\ell(\ell+d-3)}{r^2} + \frac{(d-2)(d-4)f(r)}{4r^2} + \frac{(d-2)f'(r)}{2r}\right). \label{vm}
\end{equation}
$\ell\in \mathbb{N}$ is the multipole number, and $\ell (\ell +d-3)$ are the eigenvalues of the laplacian on the $(d-2)$-sphere $\mathbb{S}^{d-2}$.

Since the test field is minimally coupled, $V_{\textsf{M}}(r)$ does not have explicit $\lambda'$ corrections. Once we replace $f(r)$ by its
$\lambda'$-dependent expression (\ref{1}), we can expand this effective potential considering terms up to first order in $\lambda'$, with the result
\bea
V_{\textsf{M}}(r) &=& V^0_{\textsf{M}}(r) + \lambda' V^1_{\textsf{M}}(r), \\
V^0_{\textsf{M}}(r) &=& V_{\textsf{M}} \left[ f_0(r) \right]
=f_0(r) \left(\frac{(d-4) (d-2) f_0(r)}{4 r^2}+\frac{(d-2) f_0'(r)}{2 r}+\frac{\ell (\ell +d-3)}{r^2}\right), \label{vm0} \\
V^1_{\textsf{M}}(r)&=& -\frac{(d-4) (d-3)}{4} \frac{R_0^{2d-4}}{r^{3d-2}} \left(\left(d^2 -2 \ell^2 +6 \ell -2 d (\ell+1)\right) r^d-(2 d-3) (d-2) R_0^{d-3} r^3\right). \label{vm1}
\eea

\subsection{Scalar type gravitational perturbations}
\noindent

Scalar type gravitational perturbations of (\ref{bw}) are governed by the master equation (\ref{potential0}), with the effective
potential taking the form \cite{Konoplya:2008ix}
\begin{equation}
V_{\textsf{S}}(r) = \frac{f(r) U(r)}{64 r^2 (d-3)^2 A(r)^2 q(r)^8 \left(4 c\, q(r)+(d-1) R(r) \left(q(r)^2 -1 \right) \right)^2}
\label{2}
\end{equation}
where $q(r)$ was defined in (\ref{bw}) and
\bea
A(r) &=& \frac{1}{q(r)^2}\left(\frac{1}{2} + \frac{1}{d-3}\right) + \left(\frac{1}{2} - \frac{1}{d-3}\right), \label{a} \\
R(r) &=& \frac{r^2}{\lambda' R_0^2 (d-3)(d-4)} \hspace{10pt}, \label{r} \\
c&=& \frac{\ell (\ell +d-3)}{d-2}-1. \label{c}
\eea
Finally, the function $U$ takes a rather lengthy form (see appendix \ref{appendix:A}).

Expanding the effective potential above to first order in $\lambda'$, we obtain the potential corresponding to scalar perturbations of the solution
(\ref{1}) we are looking for as
\begin{equation}
    V_{\textsf{S}}(r) = V^0_{\textsf{S}}(r) + \lambda' V^1_{\textsf{S}}(r),
\end{equation}
where $V^0_{\textsf{S}}$ is the uncorrected effective potential present in the master equation associated with scalar type gravitational
perturbations of the $d$-dimensional Tangherlini black hole, taking the form \cite{ik03a} of the corresponding minimal effective potential given by
(\ref{vm0}):
\begin{equation}
    V^0_{\textsf{S}}(r)=V^0_{\textsf{M}}(r) \label{vs0}.
\end{equation}
The expression for $V^1_{\textsf{S}}$ is rather lengthy (see appendix \ref{appendix:A}).

\subsection{Vector type gravitational perturbations}
\noindent

Vector type gravitational perturbations of (\ref{bw}) are governed by the master equation (\ref{potential0}), with the effective
potential taking the form \cite{Gleiser:2005ra}
\begin{equation}
    V_{\textsf{V}}(r) = f(r)\left[\frac{(d-2)c}{r^2}A(r) + K(r)\left(\frac{d^2K}{dr^2}(r) + \frac{df}{dr}(r)\frac{dK}{dr}(r)\right)\right]
\end{equation}
where
\begin{equation}
    K(r) = \frac{1}{\sqrt{r^{d-2} A(r) q(r)}},
\end{equation}
$A(r)$ is given by (\ref{a}) and $q(r)$ was defined in (\ref{bw}). Since we are only considering terms up to first order in $\lambda'$, we expand
the effective potential above as
\begin{equation}
  V_{\textsf{V}}(r) = V^0_{\textsf{V}}(r) + \lambda' V^1_{\textsf{V}}(r)
\end{equation}
where $V^0_{\textsf{V}}$ is the uncorrected effective potential present in the master equation associated with vector type gravitational perturbations of the $d$-dimensional Tangherlini black hole, taking the form \cite{ik03a}
\begin{equation}
    V^0_{\textsf{V}}(r) = f_0(r)\frac{ 4 (\ell-1) (d+\ell-2)+ (d-2) \left(d- 3 (d-2) \left(\frac{R_0}{r}\right)^{d-3}\right)}{4 r^2} \label{vv0}
\end{equation}
and
\begin{equation}
\begin{split}
    V^1_{\textsf{V}}(r) = -\frac{R_0^{d-1}}{r^{3 d+1}} \frac{(d-4)}{2} \left[- R_0^{d-3} \left((3 d-5) \ell^2+ (d-3) (3 d-5) \ell+\frac{3}{2} (d-12) (d-3) d-36\right) r^{d+3} \right. \\+2 (d-1)\left. \left(d (\ell-3)+\ell (\ell-3)+4\right) r^{2 d}+ \left(d (d (4 d-39)+91)-62\right) \frac{R_0^{2d-6}}{2} r^6\right].
    \end{split} \label{vv1}
\end{equation}

\subsection{Tensor type gravitational perturbations}
\noindent

Tensor type gravitational perturbations of (\ref{bw}) are governed by the master equation (\ref{potential0}), with the effective
potential taking the form \cite{Dotti:2005sq}
\begin{equation}
    V_{\textsf{T}}(r) = f(r)\left[ \frac{\ell(\ell+d-3)}{r^2}\left(3 - \frac{B(r)}{A(r)}\right) + K(r)\left(\frac{d^2K}{dr^2}(r) + \frac{df}{dr}(r)\frac{dK}{dr}(r)\right)\right]
\end{equation}
where
\begin{equation}
    B(r) = A(r)^2\left(1 + \frac{1}{d-4}\right) + \left( 1 - \frac{1}{d-4}\right).
\end{equation}

Since we are only considering terms up to first order in $\lambda'$, we expand the effective potential above as
\begin{equation}
    V_{\textsf{T}}(r) = V^0_{\textsf{T}}(r) + \lambda' V^1_{\textsf{T}}(r).
\end{equation}
We obtain for $V^0_{\textsf{T}}(r)$ the same $\lambda'=0$ potential as for minimally coupled test fields, given by (\ref{vm0}),
and for scalar perturbations, given by (\ref{vs0}):
\begin{equation}
    V^0_{\textsf{T}}(r) =  V^0_{\textsf{S}}(r) = V^0_{\textsf{M}}(r).\label{vt0}
\end{equation}
For $V^1_{\textsf{T}}(r)$ we obtain
\begin{equation}
\begin{split}
    V^1_{\textsf{T}}(r) = \frac{R_0^{d-1}}{2r^{3 d+1}} \left(R_0^{d-3} \left( ((d-11) d+16) \ell^2+ d (d-7)^2 \ell-\frac{d}{2} ((d-7)  (d-6) d-20) \right.
   -48 \ell+8\right) r^{d+3}\\
     \left. +4 (d-1) (d (\ell-1)+(\ell-3) \ell+4) r^{2 d}+(d-4) ((d-5) d (2 d-7)-22) \frac{R_0^{2d-6}}{2} r^6\right). \label{vt1}
    \end{split}
\end{equation}

\section{Quasinormal modes, the asymptotic limit and the monodromy method}
\label{qnm}
\noindent

Quasinormal modes associated to gravitational perturbations of asymptotically flat spherically symmetric black holes are solutions to the corresponding master equation (\ref{potential0}) subject to the boundary conditions
\bea
\psi \propto e^{-i\omega x} \hspace{3pt} , \hspace{3pt} x \to +\infty \hspace{3pt} \left(r \to +\infty\right);
\label{bcx1} \\
\psi \propto e^{i\omega x} \hspace{3pt} , \hspace{3pt} x \to -\infty \hspace{3pt} \left(r \to R_H\right).
\label{bcx2}
\eea
These boundary conditions state that waves can escape into infinity or inside the black hole. They reflect the fact that black holes are dissipative systems; hence, quasinormal frequencies are complex. The time dependence of the form (\ref{time}) requires the imaginary part of the quasinormal frequencies to be positive - otherwise the perturbation would grow indefinitely with time, which would mean an instability of the black hole solution. Thus, imposing boundary conditions (\ref{bcx1}) and (\ref{bcx2}) is equivalent to the existence of two terms of the form
$e^{\pm i\omega x}.$ For $x \to +\infty$ one of such terms is exponentially growing, while the other is exponentially vanishing. The same is true for $x \to -\infty$, but with the two terms switching their behavior. This poses an operational problem in handling and distinguishing these two terms.

But one can allow $r$ (and $x$) to take complex values and consequently assume an analytic continuation of functions of $r$ to the complex plane. In this case, near the event horizon we can distinguish the two exponential terms by computing the respective monodromies around it: indeed, the boundary condition (\ref{bcx2}) can be set as a monodromy condition. Furthermore, one can take the contour of a Stokes line defined by $\Im\left(\omega x\right)=0$ in the complex $r$ plane. Through Stokes lines we have $|e^{\pm i\omega x}| = 1$: the asymptotic behavior of $e^{\pm i\omega x}$ is always oscillatory and there will be no problems with exponentially growing versus exponentially vanishing terms. Thus if one considers the Stokes lines, imposing the boundary condition (\ref{bcx1}) in the complex $r$ plane no longer poses a challenge to an approximate analytical method.

In order to compute analytical expressions for quasinormal frequencies, we resort to the monodromy method \cite{Motl:2003cd}. This method has been used for the calculation of highly damped quasinormal modes in four and higher dimensions in Einstein gravity \cite{Birmingham:2003rf,Natario:2004jd}. Corrections to these results in terms of the inverse of the overtone number have been computed analytically, using the same method, in \cite{Musiri:2003bv}. These studies have been extended to string-theoretical \cite{Moura:2021nuh}, loop quantum-corrected \cite{Babb:2011ga} and regular black holes \cite{Lan:2022qbb}.

In order to apply this method, as we mentioned, we need to consider the coordinate $r$ as complex-valued, and the master equations (\ref{potential0}) as differential equations defined on the complex plane. The general idea of the method is to pick two closed homotopic contours on the complex $r$-plane. Both these contours enclose only the physical horizon $r=R_H$: none of them encloses the origin of the complex $r$-plane nor any other complex root of the metric function $f(r)$ (``fictitious horizons''). One of these contours, the ``big contour'', seeks to encode information of the boundary condition (\ref{bcx1}) on the monodromy of $\psi$ associated with a full loop around it. The other contour, the ``small contour'', seeks to encode information of the boundary condition (\ref{bcx2}) on the monodromy of $\psi$ associated with a full loop around it. Since both contours are homotopic, the monodromy theorem asserts that the respective monodromies must be the same. Thus, equating them yields an analytic solution to the values of the quasinormal frequencies $\omega$.

We restrict our analysis in this article to the highly damped regime of quasinormal modes defined by the condition
\be
\Im\left(\omega\right) \gg \Re\left(\omega\right). \label{asy}
\ee
This condition is equivalent to $\omega$ being approximately imaginary. The definition of a Stokes line comes thus as
\begin{equation}
    \Im\left(\omega x\right) = 0 \Rightarrow \Re\left(x\right) =0.
\end{equation}

\subsection{Choice of tortoise coordinate, Stokes lines and perturbation theory}
\noindent

The tortoise coordinate (\ref{tort}) for the full Gauss-Bonnet metric is hard to deal with, as it is impossible to express it explicitly because of the complicated form of the metric function (\ref{bw}). Since we are considering the limit in which $\lambda'$ is a perturbative parameter, we can take the metric function (\ref{1}) we are working with. The corresponding tortoise coordinate $x$ can be explicitly computed: it is given, to first order in $\lambda'$, in terms of the Gauss hypergeometric function $_{2}F_{1}$, up to an integration constant, by
\bea
x&=&_{2}F_{1}\left(1,-\frac{1}{d-3};\frac{d-4}{d-3};\left(\frac{R_0}{r}\right)^{d-3}\right) r \nonumber \\
&-& _{2}F_{1}\left(2,\frac{2d-5}{d-3};1+\frac{2d-5}{d-3};\left(\frac{R_0}{r}\right)^{d-3}\right) \frac{(d-3)(d-4)}{2 (2d-5)} \left(\frac{R_0}{r}\right)^{2d-6} r \lambda' + C_X. \label{xint}
\eea
Close to the origin of the complex plane, and with an adequate choice of the constant $C_X$, this coordinate $x$ can be approximated by
\be
x \sim - \frac{1}{d-2}\frac{r^{d-2}}{R_0^{d-3}}+\frac{(d-3)(d-4)}{2} \lambda' \frac{R_0^2}{r}. \label{x0}
\ee
We see that, because of the $\lambda'$ correction, there is a singularity in the coordinate $x$ at $r=0$.

In order to apply the monodromy method, it is essential to know the topology of the Stokes lines, given by the condition $\Re(x)=0$. Indeed, the monodromy method is very sensitive to the structure of the tortoise at the origin (namely the topology of the Stokes lines), as well as to the location in the complex plane of the metric singularities ($r=0$) and branch points of the tortoise coordinate (the real and fictitious horizons).

Because of the singularity of $x$ in (\ref{x0}), close to the origin those lines $\Re(x)=0$ are very difficult to handle; one expects them in this region to look very different from the Stokes lines given by the condition $\Re(z)=0$, corresponding to the tortoise coordinate $z$ of the Tangherlini solution. This coordinate is given by
\begin{equation}
dz = \frac{dr}{f_0(r)} \label{tort2}
\end{equation}
with $f_0(r)$ defined in (\ref{deltf}). This definition of $z$ corresponds to the $\lambda'=0$ limit of the perturbative expansion (\ref{fpert}); after integration, the explicit result corresponds to the same limit in (\ref{xint}) and (\ref{x0}), without any singularity at $r=0$.

The change in the topology of the Stokes lines of the tortoise coordinate induced by the continuous variation of a parameter is not a new
phenomenon. For instance, as it is well known, both Schwarzschild--Tangherlini and extremal Reissner-Nordstr\"om black holes are two limits of nonextremal Reissner-Nordstr\"om black holes in $d$ dimensions, corresponding to variations of the black hole charge $Q$ respectively from 0 to the extremal charge-to-mass bound ($Q=M$ in adequate units). In both these cases, the quasinormal frequencies cannot be obtained by taking the corresponding limits of the quasinormal frequencies for the non–extremal solution. In the first case, discussed in \cite{Andersson:2003fh}, since the metric function is distinct, the topology of the Stokes lines corresponding to Reissner-Nordstr\"om black holes is different from the one of Schwarzschild--Tangherlini black holes (in this sense, the Schwarzschild limit is singular). For extremal and nonextremal Reissner-Nordstr\"om black holes in $d$ dimensions the topology of the Stokes lines is the same, but the number of real and fictitious horizons is obviously different. As discussed in \cite{Natario:2004jd}, in the nonextremal case the ``big'' monodromy contour passes between the two physical horizons. It turns out that one cannot take the extremal limit without crossing the inner horizon through the ``big'' monodromy contour, thus invalidating taking the limit of the monodromy.

In general, when taking limits of the parameters with the monodromy method, one must always check whether one is crossing singularities/branch points or changing the topology of the contour. In an affirmative case, the limit on the parameters will not be valid and one will have to address the calculation separately.

The case we are dealing with in this work has a different nature: it corresponds to introducing a new parameter $\lambda'$, turning it from 0 to nonzero. The number of horizons (physical and fictitious) remains unchanged after introducing such parameter. Differently than the electric charge in the Reissner-Nordstr\"om solution, this parameter $\lambda'$ is always infinitesimal (actually, perturbative). From (\ref{deltf}) the corrections $\delta f$ to the metric function of the Tangherlini solution proportional to this parameter vary as $\left(\frac{R_0}{r}\right)^{2d-4}$, which means they decrease with $r$ faster than the uncorrected metric function $f_0$ itself. Far from the origin, the effects of $\lambda'$ are expected to  become less significant: the tortoise coordinate $x$ should be very well approximated by $z$, and the respective Stokes lines should be very similar.

\begin{figure}[H]
\centering
\includegraphics[width=0.4\textwidth]{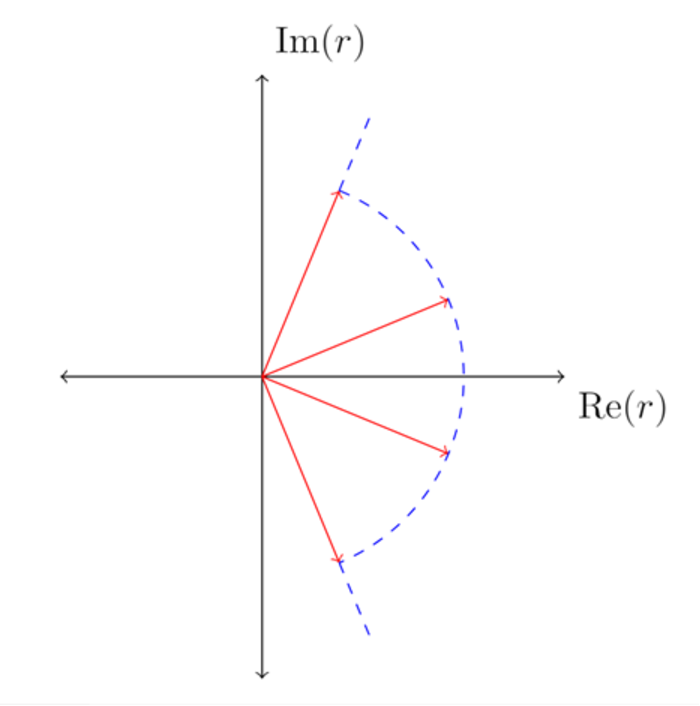}
\caption{Schematic representation of the small arc-shaped portion of the big contour, in the surroundings of $r = 0$, as the blue dashed line. The Stokes lines corresponding to the coordinate $z$ are represented by red curves. Naturally, not all Stokes lines are depicted.}
\label{fig1}
\end{figure}

Close to the origin, the effects of $\lambda'$ can be significant, as one can see for instance from the simple pole at $r=0$ of (\ref{x0}). In a neighborhood of the origin (and since it is a singular point) the big contour actually goes around without crossing it, as depicted in figure \ref{fig1}. In such neighborhood we can deform the Stokes lines corresponding to $x$ into the ones of $z$ without crossing any (real or fictitious) horizon. For the same approximation of $x$ by $z$ to be valid, from (\ref{x0}) we must have $\frac{(d-3)(d-4)}{2} \lambda' \frac{R_0^2}{r} \ll \frac{1}{d-2}\frac{r^{d-2}}{R_0^{d-3}}$ or, equivalently,
\be
r \gg \sqrt[d-1]{\frac{(d-2)(d-3)(d-4)}{2} \lambda'} R_0. \label{cons}
\ee
This condition must be verified in order to use $z$ as an approximation to $x$. It was obtained from (\ref{x0}), assuming $r$ to be close to the origin. One may wonder if this assumption of smallness of $r$ is consistent with the result (\ref{cons}). These two simultaneous assumptions are essential for the consistency of our approximation: namely both should be obeyed by the radius of the arc-shaped portion of the big contour depicted in figure \ref{fig1}. Indeed, in order to compute the monodromy associated to the big contour, we will need to solve the master equation near the origin, but that same arc-shaped portion should begin and end at Stokes lines, which should be similar to those of $z$.

The answer is positive: since $\lambda'$ is a perturbative infinitesimal parameter, $r$ can verify the condition (\ref{cons}) and still be ``small''. Therefore we do not need to worry about the Stokes lines corresponding to $x$ and we can take as the big contour the same one taken in Einstein gravity. The $\lambda'$ terms will be treated as higher derivative corrections, according to our perturbative approach. A similar perturbative approach for the calculation of asymptotic quasinormal modes through the monodromy method has also been taken, in a different context (next to leading order terms in the mode number), in the work \cite{Cardoso:2003vt}.

Taking the second order derivative in (\ref{potential0}) with respect to $z$, we can rewrite the master equations in terms of this coordinate as

\begin{equation}
    \left(\frac{dz}{dx}\right)^2\frac{d^2\psi}{dz^2} + \frac{d}{dr}\left(\frac{dz}{dx}\right)\frac{dr}{dx}\frac{d\psi}{dz} + \left(\omega^2 - V_a\right)\psi = 0.
    \label{potentialp}
\end{equation}
with $a=\textsf{S},\textsf{V},\textsf{T},\textsf{M}$. Also from (\ref{fpert}) and the definitions (\ref{tort}), (\ref{tort2}) we  can write, to first order in $\lambda'$,
\begin{equation}
    \left(\frac{dz}{dx}\right)^2 = 1 + 2\lambda' \delta f(r)
\end{equation}
\begin{equation}
    \frac{d}{dr}\left(\frac{dz}{dx}\right)\frac{dr}{dx} = \lambda' \frac{d (\delta f)}{dr}(r) .
\end{equation}

Expanding the solution of (\ref{potentialp}) as
\begin{equation}
    \psi(r) = \psi_0(r) + \lambda' \psi_1(r),
\end{equation}
replacing the expansion above in (\ref{potentialp}) and solving the equation perturbatively in powers of $\lambda'$, we obtain two distinct linear ordinary differential equations. The first one, homogeneous, of zeroth order in $\lambda'$, is
\begin{equation}
    \frac{d^2\psi_0}{dz^2}+ \left(\omega^2 - V_a^0\right)\psi_0 = 0.
    \label{potentialz0}
\end{equation}
The second one, nonhomogeneous, of first order in $\lambda'$, is
\begin{equation}
    \frac{d^2\psi_1}{dz^2}+ \left(\omega^2 - V_a^0\right)\psi_1 = \xi_a
    \label{potentialz1}
\end{equation}
where
\begin{equation}
    \xi_a = \xi_1 \frac{d^2\psi_0}{dz^2} + \xi_2 \frac{d\psi_0}{dz} + (\xi_a)_3\psi_0. \label{xi}
\end{equation}
In the expression above, we defined functions
\begin{equation}
    \xi_1(r) =  -2\delta f(r) \label{xi1}
\end{equation}
\begin{equation}
    \xi_2(r) = -\frac{d(\delta f)}{dr}(r)f(r) \label{xi2}
\end{equation}
\begin{equation}
    (\xi_a)_3(r)= V_a^1(r). \label{xi3}
\end{equation}
The subscript $a$ reflects the dependence of $\xi_a$ on the kind of potential that one is considering (gravitational perturbation or scalar field), a dependence which is explicit through $(\xi_a)_3$ but, from (\ref{potentialz0}), also implicit in $\psi_0$.

\subsection{Behavior close to the origin}
\noindent

In the sequence, namely for the differential equations (\ref{potentialz0}), (\ref{potentialz1}) and the functions $(\xi_a)_3$ in (\ref{xi3}),
we will need the asymptotic behavior near the origin of the complex $r$-plane of $\xi_1, \xi_2$ and each of the effective
potentials. We will also need to rewrite all these asymptotic expansions with respect to the tortoise coordinate $z$ of $d$-dimensional Tangherlini
black hole defined in (\ref{tort2}). This coordinate is given by the $\lambda'=0$ limit of (\ref{xint}), which can be rewritten in the form
\begin{equation}
    z(r) = r + \frac{1}{2}\sum_{n=0}^{d-4}\frac{1}{k_n}\log\left(1-\frac{r}{R_n}\right)
\end{equation}
where
\begin{equation}
    k_n = \frac{1}{2}f_0'(R_n) \hspace{10pt} ; \hspace{10pt} R_n = R_0 e^{2\pi i \frac{n}{d-3}} \label{hors}
\end{equation}
for $0 \le n \le d-4$.
We see that $z(r)$ is a multivalued function: from (\ref{tort2}), each zero of $f_0(r)$ is a branch point. There are $d-3$ zeros of $f_0(r)$, corresponding to the values of $R_n$. Of these, only the solution $R_0=R_H$ corresponds to a physical horizon; the other $d-4$ zeros are the ``fictitious'' horizons.

Near the origin of the complex $r$-plane we have the $\lambda'=0$ limit of (\ref{x0}):
\begin{equation}
    z(r) \sim - \frac{r^{d-2}}{(d-2) R_0^{d-3}}. \label{z0}
\end{equation}

The asymptotic behaviors for the effective potentials can be obtained after some algebraic manipulations.

For the potentials (\ref{vm0}) and (\ref{vm1}) for minimally coupled scalar fields, we get
\bea
V^0_{\textsf{M}}(r) &\sim& -\frac{(d-2)^2 R_0^{2d-6}}{4 r^{2d-4}}, \label{vm0r} \\
V^1_{\textsf{M}}(r) &\sim& \frac{(d-4) (d-3) (2 d-3) (d-2)}{4} \frac{R_0^{3d-7}}{r^{3 d-5}}. \label{vm1r}
\eea
For the potentials (\ref{vs0}) and (\ref{vs1}) for scalar perturbations,
\bea
V^0_{\textsf{S}}(r) &\sim& -\frac{(d-2)^2 R_0^{2d-6}}{4 r^{2d-4}}, \label{vs0r} \\
V^1_{\textsf{S}}(r) &\sim& \frac{(d-4) ((d-5) d (2 d-7)-22)}{4} \frac{R_0^{3d-7}}{r^{3 d-5}}. \label{vs1r}
\eea
For the potentials (\ref{vv0}) and (\ref{vv1}) for vectorial perturbations,
\bea
V^0_{\textsf{V}}(r) &\sim& \frac{3 (d-2)^2 R_0^{2d-6}}{4 r^{2d-4}}, \label{vv0r} \\
V^1_{\textsf{V}}(r) &\sim& -\frac{(d-4) (d (d (4 d-39)+91)-62) }{4} \frac{R_0^{3d-7}}{r^{3 d-5}}. \label{vv1r}
\eea
For the potentials (\ref{vt0}) and (\ref{vt1}) for tensorial perturbations,
\bea
V^0_{\textsf{T}}(r) &\sim& -\frac{(d-2)^2 R_0^{2d-6}}{4 r^{2d-4}}, \label{vt0r} \\
V^1_{\textsf{T}}(r) &\sim& \frac{(d-4) ((d-5) d (2 d-7)-22)}{4} \frac{R_0^{3d-7}}{r^{3 d-5}}. \label{vt1r}
\eea
We knew that $V^0_{\textsf{M}} = V^0_{\textsf{T}} = V^0_{\textsf{S}}$; it is not a surprise that their asymptotic behaviors are also the same. From
(\ref{vs1r}) and (\ref{vt1r}) we immediately notice that asymptotically we also have $V^1_{\textsf{T}} = V^1_{\textsf{S}}.$

Finally, replacing (\ref{z0}) in the asymptotic expressions for the potentials above yields
\be
V^0_{\textsf{M}}(z) = V^0_{\textsf{S}}(z) = V^0_{\textsf{T}}(z) \sim -\frac{1}{4z^2}. \label{vst0z}
\ee
For minimally coupled fields we also have
\begin{equation}
V^1_{\textsf{M}}(z) \sim \frac{(d-4) (d-3) (2 d-3) (d-2)}{4} \frac{R_0^{\frac{d-1}{d-2}}}{(-(d-2) z)^{\frac{3d-5}{d-2}}}, \label{vm1z}
\end{equation}
while, for scalar and tensorial perturbations,
\be
V^1_{\textsf{S}}(z) \sim V^1_{\textsf{T}}(z) \sim \frac{(d-4) ((d-5) d (2 d-7)-22)}{4} \frac{R_0^{\frac{d-1}{d-2}}}{(-(d-2) z)^{\frac{3d-5}{d-2}}}. \label{vst1z}
\ee
For vectorial perturbations the corresponding results are
\bea
V^0_{\textsf{V}}(z) &\sim& \frac{3}{4z^2}, \label{vv0z} \\
V^1_{\textsf{V}}(z) &\sim& -\frac{(d-4) (d (d (4 d-39)+91)-62) }{4} \frac{R_0^{\frac{d-1}{d-2}}}{(-(d-2)z)^{\frac{3d-5}{d-2}}}. \label{vv1z}
\eea

Near the origin of the complex $r$-plane, from (\ref{deltf}), (\ref{xi1}), (\ref{xi2}) we can write the asymptotic expansions for $\xi_1, \xi_2$:
\bea
\xi_1(r) &\sim& (d-4)(d-3) \frac{R_0^{d-1}}{r^{d-1}}, \label{xi10} \\
\xi_2(r) &\sim& \frac{(d-4) (d-3) (d-1)}{2} \frac{R_0^{2d-4}}{r^{2d-3}}. \label{xi20}
\eea
Rewriting these expressions with respect to the tortoise coordinate $z$, given by (\ref{z0}), yields
\begin{equation}
\xi_1(z) \sim (d-4) (d-3) \frac{R_0^{\frac{d-1}{d-2}}}{(-(d-2) z)^{\frac{d-1}{d-2}}}
\end{equation}
and
\begin{equation}
\xi_2(z) \sim \frac{(d-4) (d-3) (d-1)}{2} \frac{R_0^{\frac{d-1}{d-2}}}{(-(d-2)z)^{\frac{2d-3}{d-2}}}.
\end{equation}

Finally, from (\ref{xi3}), the asymptotic expansions for $(\xi_a)_3$ for $a = \textsf{M, T, V, S}$ can be obtained from the corresponding expansions of $V_a^1$ given by (\ref{vm1z}), (\ref{vst1z}) and (\ref{vv1z}).

\subsection{The monodromy of the big contour}
\label{qnm3}

\begin{figure}[h]
\centering
\includegraphics[width=0.5\textwidth]{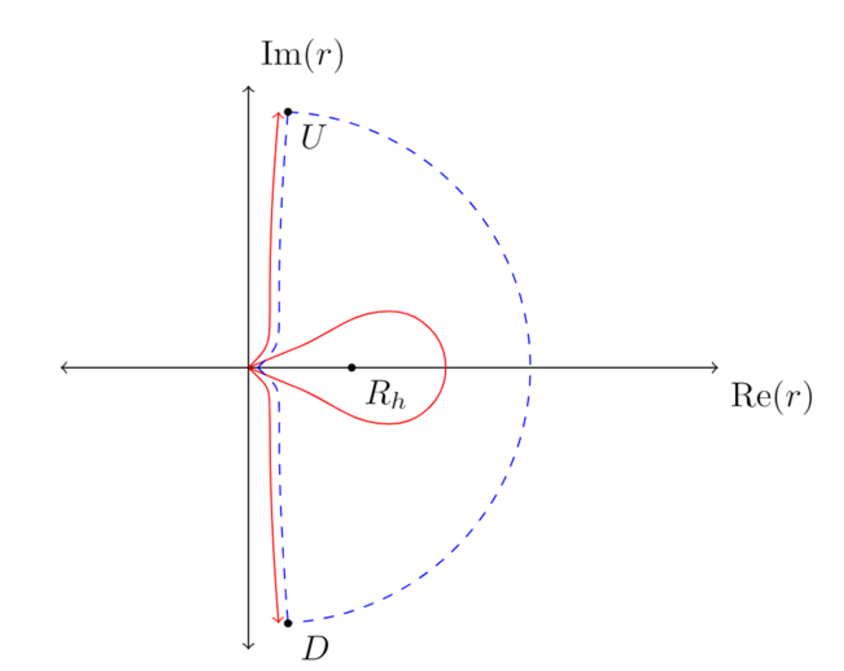}
\caption{Schematic depiction of the big contour as the blue dashed line. Red curves are Stokes lines (not all Stokes lines are depicted). $D$ and $U$ are the regions where the boundary condition (\ref{bcx1}) may be imposed.}
\label{fig2}
\end{figure}

Now we compute the monodromy of $\psi$ associated with a full clockwise loop around the big contour, depicted in figure (\ref{fig2}), starting and ending at $D$.

In order to attain reliable information on how $\psi$ changes around the big contour, we resort to WKB theory. Near the origin of the complex $r$-plane, the WKB approximation fails as $r=0$ is a singular point of $V_a^0$ for every $a=\textsf{S},\textsf{V},\textsf{T},\textsf{M}$. Thus, we need to analytically solve the differential equations (\ref{potentialz0}) and (\ref{potentialz1}) in this region.

From the asymptotic limits (\ref{vst0z}), (\ref{vv0z}), equation (\ref{potentialz0}) can be written as
\begin{equation}
    \frac{d^2\psi_0}{dz^2} + \left(\omega^2 - \frac{j^2-1}{4z^2}\right)\psi_0 = 0, \label{psi0e}
\end{equation}
with $j =0$ for the potentials in (\ref{vst0z}), and $j=2$ for the potential in (\ref{vv0z}). Following the procedure of \cite{Motl:2003cd,Natario:2004jd} we considered the general solution, for arbitrary $j$, of the above differential equation, and at the end taking the adequate limit $j \rightarrow 0$ or $j \rightarrow 2$. Such solution is given by
\begin{equation}
    \psi_0(z) = A_+ \sqrt{2\pi} \sqrt{\omega z} J_{\frac{j}{2}}(\omega z) + A_- \sqrt{2\pi} \sqrt{\omega z} J_{-\frac{j}{2}}(\omega z), \label{psi0}
\end{equation}
where $J_{\pm\frac{j}{2}}(\omega z)$ are Bessel functions of the first kind and $A_+,A_-$ arbitrary constants.

From here we can proceed in an identical fashion to the analogous computation performed in \cite{Moura:2021nuh}. The final monodromies are given
in terms of quantities that depend (in a complicated way) on the type of perturbation through the perturbation potentials $V_a$ and the value of $j$. We have
\begin{equation}
    \left(\mathcal{M}_a\right)_1 = \left(\frac{A_+e^{5i\alpha_+} + A_-e^{5i\alpha_-}}{A_+e^{i\alpha_+} + A_-e^{i\alpha_-}}\right) \left(1+ \lambda' \delta\left(\mathcal{M}_a\right)_1\right)e^{-i\omega  \Delta_z}
    \label{14}
\end{equation}
with
\bea
&&\Delta_z\ = -\frac{2\pi i R_0}{d-3}, \label{16}\\
&&\alpha_{\pm} = \frac{\pi}{4}\left(1 \pm j\right), \label{alfa}\\
&&A_{\pm}(j) =\mp \frac{i e^{-i\alpha_{\mp}}}{2} \csc\left(\frac{j \pi}{2}\right). \label{a0s}
\eea
Taking both the limit $j\to 2$ (for vectorial perturbations) and $j\to 0$ (for the other cases), we get in both cases the same result for the uncorrected part of all monodromies \cite{Natario:2004jd}:
\begin{equation}
\frac{A_+e^{i5\alpha_+} + A_-e^{i5\alpha_-}}{A_+e^{i\alpha_+} + A_-e^{i\alpha_-}} = -3. \label{m3}
\end{equation}
The correction term to the monodromies in (\ref{14}) is given by
\begin{equation}
\delta \left(\mathcal{M}_a\right)_1 = \frac{\Lambda_F^+e^{i5\alpha_+} +\Lambda_F^-e^{i5\alpha_-}}{A_+ e^{5i\alpha_+}+ A_-e^{5i\alpha_-}} - \frac{\Lambda_I^+e^{i\alpha_+} +\Lambda_I^-e^{i\alpha_-}}{A_+e^{i\alpha_+} + A_-e^{i\alpha_-}}\label{dm}
\end{equation}
with
\bea
&&\Lambda_I^\pm(d,j,\omega,R_0) = \left(R_0 \omega \right)^{\frac{d-1}{d-2}} \sum_{k=1}^3\Theta_k^\pm(d,j,1,1), \label{lambi}\\
&&B_{\pm}(d,j,\omega,R_0) =A_{\pm} \left(\Lambda_I^+e^{-i\alpha_+}+\Lambda_I^-e^{-i\alpha_-}\right), \label{b0s}\\
&&\Lambda^\pm_F(d,j,\omega,R_0) = \left(R_0 \omega \right)^{\frac{d-1}{d-2}} \sum_{k=1}^3 \Xi_k^\pm(d,j,1,1) + B_\pm. \label{fi23pi}
\eea
Explicit expressions for the functions $\Theta_k^\pm, \Xi_k^\pm, \, k=1, 2, 3$ can be found in \cite{Moura:2021nuh}.

The sums in $k$ in the definitions of $\Lambda_I^\pm, \Lambda^\pm_F$ have their origin in the analogous sum in the definition of $\xi_a$: each of the functions $\Theta_k^\pm, \Xi_k^\pm, \, k=1, 2, 3$ in (\ref{lambi}) and (\ref{fi23pi}) corresponds to the term including $\xi_k$ in (\ref{xi}), with $\xi_k$ given in (\ref{xi1}), (\ref{xi2}), (\ref{xi3}). Each of these functions $\Theta_k^\pm, \Xi_k^\pm$ has a different form, because of the different coefficients multiplying each $\xi_k$ in (\ref{xi}), but always involving sums of products/quotients of Gamma functions, which we designate by $\mathcal{H}(m,n,k)$, given by
\begin{equation}
    \mathcal{H}(m,n,k):= \frac{\Gamma \left(\frac{1}{2}-\frac{k }{2}\right) \Gamma \left(-\frac{k }{2}\right) \Gamma \left(\frac{k }{2}+\frac{m }{2}+\frac{n}{2}+\frac{1}{2}\right)}{2 \sqrt{\pi} \Gamma \left(-\frac{k}{2}+\frac{m }{2}-\frac{n }{2}+\frac{1}{2}\right) \Gamma \left(-\frac{k }{2}+\frac{n}{2}-\frac{m }{2}+\frac{1}{2}\right) \Gamma \left(-\frac{k}{2}+\frac{m}{2}+\frac{n }{2}+\frac{1}{2}\right)}. \label{hkmn}
\end{equation}
The complete expression for $\delta \left(\mathcal{M}_a\right)_1$ in (\ref{dm}) is a sum of 96 of these terms, all with different coefficients. Specifically there are terms depending on
\be
\mathcal{H}\left(\pm \frac{j}{2}, \pm \frac{j}{2},-\frac{1}{d-2}\right), \mathcal{H}\left(\pm \frac{j}{2}, \pm \frac{j}{2},-\frac{d-1}{d-2}\right), \mathcal{H}\left(\pm \frac{j}{2}, \pm \frac{j}{2},-\frac{2d-3}{d-2}\right). \label{hj}
\ee
We will not reproduce these terms here, since they can all be found in \cite{Moura:2021nuh}. The calculation of all these expressions for a generic value of $j$ is evidently a formidably complicated task.

For the case $j=0$ (one of the two cases we will need), from (\ref{hj}) the functions $\mathcal{H}(m,n,k)$ that arise are always of the form $\mathcal{H}\left(0, 0,k\right)$. Using the known properties of the Gamma functions
\bea
&&\Gamma(x+1) = x \Gamma(x), \\\label{gfp1}
&&\Gamma(x) \Gamma(1-x) = \frac{\pi}{\sin(\pi x)}, \label{gfp2}
\eea
one can easily show that these functions are given by
\be
\mathcal{H}\left(0, 0,k\right)= \frac{\sqrt{\pi} \Gamma\left(-\frac{k}{2}\right)}{2 \Gamma\left(\frac{1-k}{2}\right)^3
\sin\left(\frac{1-k}{2} \pi\right)}. \label{h0}
\ee
One can then evaluate $\mathcal{H}\left(0, 0,k\right)$ for the values of $k$ referred in (\ref{hj}).

The dependence of the monodromy $\left(\mathcal{M}_a\right)_1$ on the type of perturbation is expressed in the value of $j$, as we mentioned (or equivalently on the potential $V^0_a$), but also on $(\xi_a)_3$ (or equivalently on the potential $V^1_a$). The result for $\delta\left(\mathcal{M}_a\right)_1$ in (\ref{dm}) is always of the form
\begin{equation}
    \delta\left(\mathcal{M}_a\right)_1 = \left(\frac{R_0 \omega}{d-2}\right)^{\frac{d-1}{d-2}} e^{-\frac{2i\pi}{d-2}} \Pi_a,
    \label{19}
\end{equation}
with a factor $\Pi_a$ depending on the type of perturbation.

For test scalar fields, with $j=0$ and $(\xi_{\textsf{M}})_3$ given by (\ref{vm1z}), we obtained
\begin{equation}
    \Pi_{\textsf{M}} = \frac{8}{3} \frac{\pi^2}{2^{\frac{1}{d-2}}} \frac{(d-4) (d-3) (d-2)}{d-1} \frac{\Gamma \left(\frac{1}{d-2}\right) }{ \left[\Gamma \left(\frac{d-1}{2d-4}\right)\right]^4} \sin \left(\frac{\pi }{2(d-2)}\right). \label{pim}
\end{equation}

For tensorial perturbations, we also take $j=0$ but considering $(\xi_{\textsf{T}})_3$ given by (\ref{vst1z}). For scalar perturbations we should consider $j=0$ and $(\xi_{\textsf{S}})_3$ but, as we know from (\ref{vst1z}), $(\xi_{\textsf{S}})_3=(\xi_{\textsf{T}})_3$. This way, the result for
$\Pi_{\textsf{S}}$ will be the same as the one for $\Pi_{\textsf{T}}$.

For vectorial perturbations we must take $j=2$. From (\ref{hj}) this means the functions $\mathcal{H}(m,n,k)$ that arise in the calculation of $\left(\mathcal{M}_{\textsf{V}}\right)_1$ are always of the form $\mathcal{H}\left(\pm 1, \pm 1,k\right)$. From the definition (\ref{hkmn}) and using (\ref{gfp1}), one can show that
\be
\mathcal{H}\left(1, 1,k\right)= \mathcal{H}\left(-1, -1,k\right)= -\mathcal{H}\left(-1, 1,k\right)= -\mathcal{H}\left(1, -1,k\right)= \frac{1+k}{1-k} \mathcal{H}\left(0, 0,k\right) \label{h1}
\ee
with $\mathcal{H}\left(0, 0,k\right)$ given by (\ref{h0}). One can this way evaluate $\mathcal{H}\left(\pm 1, \pm 1,k\right)$ for the values of $k$ referred in (\ref{hj}). We can then proceed computing $\Pi_{\textsf{V}}$, taking $j=2$ also in the adequate coefficients and the value (\ref{vv1z}) for $(\xi_{\textsf{V}})_3$. Rather surprisingly, because of the changes motivated by taking $j=2$ in (\ref{h1}) and in the adequate coefficients, $\Pi_{\textsf{V}}$ turned out to be identical to $\Pi_{\textsf{T}}$ and $\Pi_{\textsf{S}}$, despite $(\xi_{\textsf{V}})_3 \neq (\xi_{\textsf{T}})_3, \, (\xi_{\textsf{S}})_3$:

\begin{equation}
    \Pi_{\textsf{T}}=\Pi_{\textsf{V}} = \Pi_{\textsf{S}} = \frac{2 \sqrt{\pi}}{3}\frac{(d (d - 5) + 2) (d - 4)}{d-1} \frac{\Gamma \left(\frac{1}{2 (d-2)}\right) \Gamma \left(\frac{d-3}{2 (d-2)}\right)}{\Gamma \left(\frac{d-1}{2 (d-2)}\right)^2}\sin\left(\frac{\pi}{d-2}\right). \label{pit}
\end{equation}

To summarize, the monodromy of $\psi$ associated with a full clockwise loop around the big contour is given by
\begin{equation}
\left(\mathcal{M}_a\right)_1 = - 3 \left(1+ \lambda' \left(\frac{R_0 \omega}{d-2}\right)^{\frac{d-1}{d-2}} e^{-\frac{2i\pi}{d-2}} \Pi_a \right)e^{-i\omega \Delta_z}, \label{m14}
\end{equation}
with $\Delta_z$ given by (\ref{16}) and $\Pi_a$ given by (\ref{pim}), for test scalar fields, and by (\ref{pit}), for all kinds of gravitational perturbations.

\subsection{The monodromy of the small contour}
\noindent

\begin{figure}[H]
\centering
\includegraphics[width=0.5\textwidth]{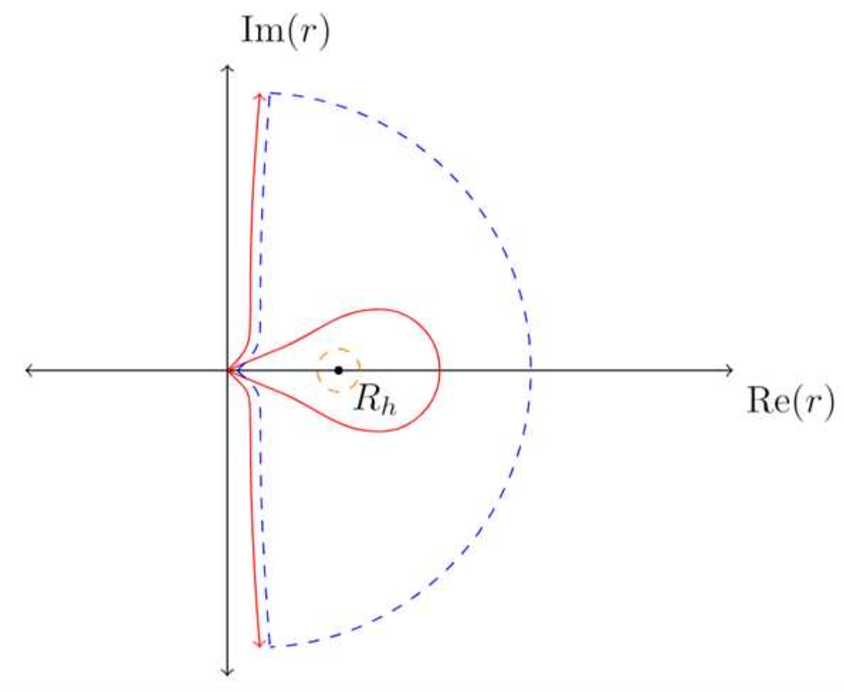}
\caption{Schematic depiction of the small and big contours as the orange and blue dashed lines respectively. The orange contour is to be interpreted as arbitrarily close to $R_H$. The Stokes lines are depicted by red curves. Naturally, not all Stokes lines are depicted.}
\label{fig3}
\end{figure}

Now we compute the monodromy of $\psi$, associated with a full clockwise loop around the small contour, depicted in figure (\ref{fig3}).

From (\ref{rh}) we write, up to first order in $\lambda'$, the asymptotic expansion near the real horizon $R_H$
\begin{equation}
    \frac{1}{f(r)} \sim \frac{1}{(d-3)\left(1-\frac{R_H}{r}\right)}\left[1 +  \lambda' \frac{(d-4) (d-1)}{ 2} \right].
\end{equation}

From the definition (\ref{tort}), integrating the expression above yields the asymptotic expansion of the tortoise coordinate $x$ near $R_H$:
\begin{equation}
x(r) \sim \frac{R_H}{d-3}\log\left(r-R_H\right)\left[1 + \lambda' \frac{(d-4)(d-1)}{2}\right]. \label{xhor}
\end{equation}
The monodromy of $x(r)$ in (\ref{xhor}), associated with a full clockwise loop around the real horizon $R_H$, is given by
\begin{equation}
    \Delta_x = -\frac{2\pi i R_H}{d-3}\left[1 +  \lambda' \frac{(d-4)(d-1)}{2}\right] = -\frac{2\pi i R_0}{d-3}\left[1 + \lambda' \frac{(d-4)(d-2)}{2}\right].
    \label{17}
\end{equation}

One easily checks that
\begin{equation}
    \lim_{r \to R_H} V_a(r) = 0
\end{equation}
for all $a=\textsf{S},\textsf{V},\textsf{T},\textsf{M}$. Thus, near the event horizon $R_H$, we can write the master equations as
\begin{equation}
    \frac{d^2\psi}{dx^2} + \omega^2\psi = 0
\end{equation}
whose solution, imposing the boundary condition (\ref{bcx2}) in the event horizon, is
\begin{equation}
    \psi(x) = \mathcal{C}_+ e^{i\omega x}.
\end{equation}
for some $\mathcal{C}_+ \in \mathbb{C}$. Computing the monodromy of the expression above, associated with a full clockwise loop around the event horizon $R_H$, yields
\begin{equation}
    (\mathcal{M}_a)_2 = e^{i \omega \Delta_x}.
    \label{m15}
\end{equation}

\subsection{Equating monodromies: the final result}
\noindent

Now, we want to relate the monodromies $(\mathcal{M}_a)_1$ and $(\mathcal{M}_a)_2$.

We notice the big contour is homotopic to the small one. This is so, because one can continuously deform the big contour into the small one. Thus, using the monodromy theorem, we know the monodromies of $\psi$, associated with the full clockwise loops around the big and the small contours, are equal. Hence, the equation
 \begin{equation}
     (\mathcal{M}_a)_1 = (\mathcal{M}_a)_2
 \end{equation}
must hold. Using equations (\ref{m14}) and (\ref{m15}), we can rewrite the equation above as
\begin{equation}
     -3(1 + \lambda' \delta (\mathcal{M}_a)_1)e^{-i\omega (\Delta_x+ \Delta_z)} = 1.
\end{equation}
Taking the logarithm on both sides of the equation above yields
\begin{equation}
    \ln\left(3\right) + (2k+1)\pi i -i\omega(\Delta_x+ \Delta_z) + \log(1+\lambda'\delta(\mathcal{M}_a)_1) = 0
\end{equation}
for $k\in \mathbb{Z}$. Taylor expanding the last logarithm in the equation above, up to first order in $\lambda'$, yields
\begin{equation}
    \ln(3) + (2k + 1)\pi i - i \omega(\Delta_x+ \Delta_z) + \lambda'\delta(\mathcal{M}_a)_1  = 0.
    \label{18}
\end{equation}
Considering the expressions for $\Delta_z, \Delta_x$ (\ref{16}) and (\ref{17}), we can write
\begin{equation}
    \Delta_x+ \Delta_z = \frac{-4\pi i R_0}{d-3}\left(1 + \lambda' \frac{(d-4)(d-2)}{4}\right).
\end{equation}
The equation above and (\ref{19}) allow us to rewrite (\ref{18}) as
\begin{equation}
\ln(3) + (2k + 1)\pi i  = \omega\frac{4\pi R_0}{d-3}\left(1 + \lambda' \frac{(d-4)(d-2)}{4}\right)- \lambda'\left(\frac{R_0 \omega}{d-2}\right)^{\frac{d-1}{d-2}} e^{-\frac{2i\pi}{d-2}} \Pi_a. \label{21}
\end{equation}
The results should be written in terms of physical variables, instead of the parameter $R_0$. It is customary to write the QNM frequencies with respect to the Hawking temperature $T_H$ of the black hole given by (\ref{temp}). We can solve (\ref{temp}) for $R_0$ as a function of $T_H$ and replace it in (\ref{21}) and in the definition (\ref{lambda}) of $\lambda'$, obtaining $\lambda'=\alpha \left(\frac{4 \pi}{d-3}\right)^2 T_H^2$. Solving for $\frac{\omega}{T_H}$ one gets
\be
\frac{\omega}{T_H} = \left[\ln(3) + (2k + 1)\pi i \right] \left[1
+ \alpha \left(\frac{4 \pi}{d-3}\right)^2 T_H^2 \left(\frac{(d-4)(d-2)}{4} + \left[\frac{d-3}{4 \pi (d-2)}\right]^{\frac{d-1}{d-2}} \left[\frac{\omega}{T_H}\right]^{\frac{1}{d-2}} e^{-\frac{2i\pi}{d-2}} \Pi_a \right) \right]. \label{wtt}
\ee
This is a transcendental equation for $\frac{\omega}{T_H}$. Recalling that we have been working in the highly damped limit (\ref{asy}), we can take $\left.\frac{\omega}{T_H}\right|_{\alpha=0} \approx (2k + 1)\pi i$ in the $\alpha$ correction in (\ref{wtt}), obtaining this way the final expression for the $\alpha$-corrected asymptotic quasinormal spectra:
\be
\frac{\omega}{T_H} = \left[\ln(3) + (2k + 1)\pi i \right] \left[1
+ \alpha \left(\frac{4 \pi}{d-3}\right)^2 T_H^2 \left(\frac{(d-4)(d-2)}{4} + \left[\frac{d-3}{d-2}\right]^{\frac{d-1}{d-2}} \left[\frac{2k + 1}{4}\right]^{\frac{1}{d-2}} \frac{\Pi_a}{4 \pi} \, \mathrm{e}^{-\frac{3 \pi i}{2(d-2)}} \right) \right], \label{wtt2}
\ee
with $\Pi_a$ given by (\ref{pim}), for test scalar fields, and by (\ref{pit}), for all kinds of gravitational perturbations. We also take $k \in \mathbb{N}$ in order to have $\Im\left(\omega\right)>0$ which, from our conventions in (\ref{time}), is a necessary condition for the stability of the black hole. \footnote{Some authors use the opposite convention, i.e. $\Im\left(\omega\right)<0$. This would be equivalent to taking the complex conjugate: in (\ref{time}) and in the boundary conditions (\ref{bcx1}) and (\ref{bcx2}). In order to express our results in this convention, one should take the complex conjugate of every calculation, namely of the monodromies $(\mathcal{M}_a)_1, \, (\mathcal{M}_a)_2$ and of the final result (\ref{wtt2}).}

This expression has the same form as the ones previously derived in \cite{Moura:2021nuh} for the black hole solution with higher derivatives from string theory obtained in \cite{cmp89}, the only difference coming from the distinct values of the Hawking temperature, giving rise to a different value of the $k-$independent correction term in (\ref{wtt2}). This term will be actually negligible in the large $k$ limit, which is the asymptotic limit.

Because of the $\mathrm{e}^{-\frac{3 \pi i}{2(d-2)}}$ term the $\alpha$ correction is complex, which means it will affect both the real and the imaginary parts of $\omega$. The real part of the asymptotic limit of $\frac{\omega}{T_H}$ is then no longer equal to the universal value $\ln(3)$: because of the $\alpha$-correction, it now also depends on the spacetime dimension $d$ and on the mode number $k$. We have evaluated numerically $\left[\frac{d-3}{d-2}\right]^{\frac{d-1}{d-2}} \frac{\Pi_a}{4 \pi}$ for the relevant values of $d$ in the context of superstring theory. This factor grows monotonically with $d$, varying from approximately 0.19 (corresponding to $d=5$) to approximately 13.5 (corresponding to $d=10$), for $\Pi_{\textsf{T}}, \, \Pi_{\textsf{V}}, \, \Pi_{\textsf{S}}$. The values of $\Pi_{\textsf{M}}$, on the same range, are slightly larger, varying from approximately 0.58 (corresponding to $d=5$) to approximately 14.48. Just for comparison, $\frac{(d-2)(d-4)}{4}$ varies between 0.75 and $12$ on the same range. In order to compare the orders of magnitude of the two correction terms in (\ref{wtt2}), one must also consider the factor $\left[\frac{2k + 1}{4}\right]^{\frac{1}{d-2}}$ multiplying the term with $\Pi_a$. For values of $k$ that are large but not very large, the two correction terms in (\ref{wtt2}) are of comparable orders of magnitude and should be considered (specially for larger $d$, considering the $\frac{1}{d-2}$ power of $2k+1$). But in general, the larger the value of $k$, the more negligible the $k$-independent term becomes. For large $k$, one considers just the $k$-dependent term, which is multiplied by the phase $\mathrm{e}^{-\frac{3 \pi i}{2(d-2)}}$ and by the $\alpha=0$ term $(2k+1)\pi i$. There will be a correction term multiplied by $\cos \frac{3 \pi}{2(d-2)}$ (always positive, affecting $\Im\left(\omega\right)$) and another correction term multiplied by $\sin \frac{3 \pi}{2(d-2)}$ (also always positive, affecting $\Re\left(\omega\right)$). The relative magnitude of these two corrections depends of course on $d$ through those trigonometric terms, but we may conclude that, for every value of $d$, the higher derivative corrections we considered increase the magnitudes of both $\Re\left(\omega\right)$ and $\Im\left(\omega\right)$.

In a perturbative expansion the correction terms should be relatively small. The perturbation parameter $\lambda'$ defined in (\ref{lambda}) is supposed to be small in magnitude in our perturbative approach, but for the $k$-dependent correction term to be small (even if multiplied by $\lambda'$), $k$ cannot be too large. As we argued in \cite{Moura:2021nuh}, consistency with the condition (\ref{asy}) defining the highly damped limit used in our derivation requires $k$ to be large enough such that the condition $(2k + 1)\pi \gg \ln(3)$ is valid, but $k$ should not be arbitrarily large.

\section{Conclusions}
\noindent

In this article, we have studied quasinormal modes of spherically symmetric $d$ dimensional black holes in Gauss-Bonnet gravity, also considered perturbatively, equivalently to having large black holes, having taken the asymptotic limit and computed them analytically, using the monodromy method. We concluded that the magnitudes of both $\Re\left(\omega\right)$ and $\Im\left(\omega\right)$ increase by considering the Gauss-Bonnet higher derivative corrections, when compared to the corresponding results in Einstein gravity. Furthermore we have obtained the same result for the three different kinds of perturbations (tensorial, vectorial and scalar), which means that the isospectrality of quasinormal modes for different kinds of perturbations in the asymptotic limit is preserved in the presence of perturbative Gauss-Bonnet corrections.

This isospectrality is a remarkable result, since as we have seen in section \ref{qnm3} the expression for the monodromy of the big contour depends explicitly on the potential corresponding to the perturbation being considered (more precisely, on its asymptotic expansion close to the origin). For vectorial perturbations, these asymptotic expansions are different than for the other kinds, both at $\alpha=0$ and to first order in $\alpha$ (the Gauss-Bonnet parameter in which we have made a perturbative expansion). This difference gives rise to different asymptotic master equations (which we also considered perturbatively in $\alpha$) close to the origin. At order $\alpha=0$, the asymptotic master equation close to the origin (\ref{psi0e}) is such that the corresponding value of $j$ for vectorial perturbations is different than for the other kinds. This affects the coefficients of many terms in the monodromy, as we mentioned. To first order in $\alpha$, the asymptotic potential corresponding to vectorial perturbations is different than for the other kinds, giving rise to different values of $(\xi_a)_3$ in (\ref{xi3}) that directly affect the monodromy. The net effect of the two changes (of $j$ and $(\xi_a)_3$) for vectorial perturbations is such that the total value of the monodromy does not change: it is the same for all kinds of gravitational perturbations, and so is the spectrum of asymptotic quasinormal modes with Gauss-Bonnet corrections, to first order in $\alpha$.

We have also considered test scalar fields propagating in the same kind of black holes. These fields satisfy the same field equation, with the same potential (and logically the same spectrum of quasinormal modes), as tensorial gravitational perturbations in Einstein gravity. In the presence of higher derivative corrections the potentials become distinct and, as we have shown, so do the spectra of quasinormal modes (although the functional form of the corrected frequencies is analogous in both cases). The isospectrality of asymptotic quasinormal modes with Gauss-Bonnet corrections we observed is, therefore, valid only for gravitational perturbations.

There is a potential conflict between the asymptotic limit of quasinormal modes and the perturbative limit of large black holes that we have taken. The imaginary parts of these frequencies grow with the mode number $k$ to infinity. Although the growth of the correction terms is more moderate than the one of the uncorrected parts, they also grow arbitrarily with $k$. But the corrections to the quasinormal mode frequencies, like any perturbative correction, are supposed to be small. This means that our result is valid for values of $k$ which are large enough for the asymptotic condition to be verified, but not arbitrarily large. The maximum allowed value of $k$ depends on the magnitudes of the Gauss-Bonnet constant (which is expected to be very small), but also of the dimensional-dependent factors of the corrections. We have evaluated numerically these factors, both for the gravitational perturbations and for the test scalar fields. We concluded that, although growing with the spacetime dimension $d$, the magnitude of these factors is perfectly compatible with a perturbative expansion for the relevant values of $d$ in the context of superstring theory.

Our results have the same form as the ones for the black hole solution with higher derivatives from string theory obtained in \cite{Moura:2021nuh}. There is no difference between perturbative Gauss-Bonnet gravity and string theory concerning asymptotic quasinormal mode frequencies for spherically symmetric large black holes, at least for the cases whose results are known in string theory (tensorial perturbations and test fields). Since the only difference in these two theories, as we mentioned, is the presence (in string theory) of the (nonconstant) dilaton field, we conclude that this field does not affect the asymptotic quasinormal mode frequencies, at least for those two cases that were previously considered. In future works we plan to study the remaining cases in string theory.

The obtained results indicate a tendency of the highly damped quasinormal mode frequencies for large black holes in Gauss-Bonnet gravity. In this limit, one cannot distinguish the spectra of QNMs associated to different gravitational perturbations. It would be interesting to extend this study to black hole solutions with higher order corrections (namely $\a^3$ corrections from string theory), in order to figure out if this isospectrality is preserved also by such corrections or if it is just a feature of Einstein and Gauss-Bonnet gravity.

\paragraph{Acknowledgements}
\noindent
This work has been supported by Funda\c c\~ao para a Ci\^encia e a Tecnologia under grants IT (UIDB/50008/2020 and UIDP/50008/2020), CAMGSD/IST-ID (UIDB/04459/2020 and UIDP/04459/2020) and projects CERN/FIS-PAR/0023/2019, 2022.08368.PTDC. Jo\~ao Rodrigues is supported by Funda\c c\~ao para a Ci\^encia e a Tecnologia through the doctoral fellowship UI/BD/151499/2021.

\appendix

\section{Effective potential functions}
\label{appendix:A}
\noindent

The function $U(r)$, present in the effective potential $V_{\textsf{S}}(r)$ associated with scalar type gravitational perturbations, reads
\bea
    U(r) &=& 5 (d-1)^6 R(r) ^2 (R(r) +1)-3 (d-1)^5 R(r)  q(r) \left(24 c (R(r) +1)+(d-1) R(r) ^2\right)\nonumber \\ &&
    +2 (d-1)^4 q(r)^2 \left(168 c^2 (R(r) +1)+24 c (d-1) R(r) ^2-(d-1) R(r) ^2 (7 d (R(r) +1)+5 R(r) -3)\right)\nonumber \\ &&
    +2 (d-1)^4 R(r)  q(r)^3 \left(c (84 d (R(r) +1)+44 R(r) -84)-184 c^2+(d-1) (d+13) R(r) ^2\right) \nonumber \\ &&
    +(d-1)^3\left(384 c^3-48 c ((3 d-5) d+2) R(r) ^2+192 c^2 \left((d-15) R(r) ^2+d-11\right)\right. \nonumber \\ &&
    \left.+(d-1) R(r) ^2 (d (7 d (R(r) +1)+106 R(r) +26)-3 (55 R(r) +7))\right)q(r)^4 \nonumber \\ &&
    +(d-1)^3R(r) \left(-64 c^2 (d-38)+(d-1) ((7 d-90) d+71) R(r) ^2\right. \nonumber \\ &&
    \left.+16 c \left(13 d^2 (R(r) +1)-2 d (81 R(r) +73)+255 R(r) +303\right)\right)q(r)^5 \nonumber \\ &&
    +4(d-1)^2\left(96 c^3 (d-7)-8 c (d-1) \left(6 d^2-74 d+145\right) R(r) ^2\right. \nonumber \\ &&
    -8 c^2 (d (11 d (R(r) +1)-34 R(r) -58)-175 R(r) +9)+ (d-1)R(r)^2 (-5 (23 R(r) +79) \nonumber \\ &&
    +d (d (7 d (R(r) +1)-89 R(r) -81)+5 (41 R(r) +57))))q(r)^6 \nonumber \\ &&
    -4(d-1)^2R(r) \left(8 c^2 (d (72-13 d)+43)+(d-1) (d (d (5 d-49)+99)-63)+R(r) ^2\right. \nonumber \\ &&
    +4 c (d (d (17 d (R(r) +1)-107 R(r) -123)-39 R(r) +121)+465 R(r) +321))q(r)^7\nonumber \\ &&
    +(d-1)\left(128 c^3 (d-9) (d-5)+32 c (d-1) (d (d (8 d-55)+9)+246) R(r) ^2\right. \nonumber \\ &&
    +64 c^2 (d-5) \left(d^2+((d-4) d+49) R(r) -3\right) \nonumber \\ &&
    \left.-(d-1) R(r) ^2 (d (d (d (45 d (R(r) +1)-452 R(r) -548)+6 (217 R(r) +393))  \right. \nonumber \\ &&
    \left. -4 (349 R(r) +997))+565 R(r) +1173)\right)q(r)^8 \nonumber \\ &&
    +(d-1)R(r) \left(-64 c^2 (d-5) (d (3 d-13)+36)+(d-1) (d (3 d (d (9 d-92)+294)-1204)+635) R(r) ^2\right. \nonumber \\ &&
    -8 c (d-5) (d (d ((d-79) R(r) +d-47)+191 R(r) +127)+31 R(r) +63))q(r)^9 \nonumber \\ &&
    +2d-5\left(64 c^3 (d-5) (d-3)+8 c (d-1) (d ((d-43) d+141)-27) R(r) ^2\right. \nonumber \\ &&
    +8 c^2 (d-5) (d ((d-18) R(r) +d-2)+77 R(r) -3)+(d-1)^2R(r) ^2(-33 (R(r) -7) \nonumber \\ &&
    +d (d (9 d (R(r) +1)-35 R(r) -59)+43 R(r) +59)))q(r)^{10} \nonumber \\ &&
    -2d-5R(r) \left(24 c^2 (d-11) (d-5) (d-3)+(d-1)^2 (d ((7 d-39) d+81)-65) R(r) ^2\right. \nonumber \\ &&
    +12 c (d-7) (d-5) (d-3) (d-1) (R(r) +1))q(r)^{11} \nonumber \\ &&
    +(d-5)^2 (d-1) R(r) ^2 q(r)^{12} (16 c ((d-9) d+26)+(d-1) (d ((d-2) R(r) +d-18)-3 R(r) +77)) \nonumber \\ &&
    +(d-5)^2 (d-3)^2 (d-1)^2 R(r) ^3 q(r)^{13}.
\eea
$R(r)$ is given by (\ref{r}), $c$ was defined in (\ref{c}) and $q(r)$ was defined in (\ref{bw}).

The expression for $V^1_{\textsf{S}}$, the first-order in $\lambda'$ correction to $V_{\textsf{S}}(r)$, is given by
\bea
       V^1_{\textsf{S}}(r) &=& -\frac{(d-4) R_0^{d-1} r^{-6 d-1}}{4 (d-2)^2 \left((\ell-1) (d+\ell-2)+(d-1) \left(\frac{d-2}{2} R_0^{d-3}\right)  r^{3-d}\right)^3}
        \nonumber\\ &&
       \left(-8 (d-2)^2 (d-1) (\ell-1)^3 (d+\ell-2)^3 \left((d-3) \ell+(d-2)^2+\ell^2\right) r^{5 d}  \right. \nonumber\\&&+
       2 (d-2) (\ell-1)^2 (d+\ell-2)^2
       \left((d-2)^2 (d (d (4 d-71)+201)-152)\right.   \nonumber\\&& +
       (d-3) (d-2) (d (19 d-103)+106) \ell \nonumber\\&&+(d (d (29 d-215)+486)-338) \ell^2+
       \left.2 (5 d-7) \ell^4+4 (d-3) (5 d-7) \ell^3\right)r^{4 d+3}(d-2) R_0^{d-3}\frac{1}{2}  \nonumber\\&&-
       2 (d-2) (\ell-1) (d+\ell-2)
       \left(-(d-2)^2 (d (d (15 d-268)+747)-554)\right. \nonumber\\&&+
       (d-3) (d-2) (d (d (15 d-242)+801)-694) \ell+
       ((d-11) d+22) (d (37 d-147)+140) \ell^2 \nonumber\\&&+4 (d-3) (d (11 d-75)+94) \ell^3+
       \left.2 (d (11 d-75)+94) \ell^4\right)r^{3 d+6} (d-2)^2 R_0^{2d-6}\frac{1}{4}  \nonumber\\&&-
       2(d-1)\left(-(d-2)^3 (d (d (7 d-60)-113)+370)+\right.
       (d-3) (d-2)^2 (d (7 (d-6) d-255)+698) \ell \nonumber\\&&+
       (d (d (d (-3 (d-48) d-1661)+7112)-12992)+8624) \ell^2 \nonumber\\&&-
       \left.4 (d-3)^2 (d (5 d-62)+108) \ell^3-2 (d-3) (d (5 d-62)+108) \ell^4\right)
       (d-2)^3 R_0^{3d-9}\frac{1}{8} r^{2 d+9} \nonumber\\&&- 2(d-1)^2\left((d-2)^2 (d (d (d+8)-131)+134)\right.-
      2 (d ((69-7 d) d-172)+116) \ell^2 \nonumber\\&& \left. +2 (d-3) (d (d (7 d-69)+172)-116) \ell\right)
        (d-2 )^4 R_0^{4d-12}\frac{1}{16} r^{d+12} \nonumber\\&&  \left.+4 (d-1)^3 ((d-5) d (2 d-7)-22)(d-2)^5 R_0^{5d-15}\frac{1}{32}  r^{15}\right)
        \label{vs1}.
\eea

\end{document}